\documentclass[pla,twocolumn,showpacs,preprintnumbers,amsmath,amssymb]{revtex4}
\usepackage{mathrsfs}
\usepackage{graphicx}

\begin{document}

\title{Competition between paramagnetism and diamagnetism in charged Fermi
gases}

\author{Xiaoling Jian}
\author{Jihong Qin }
\author{Qiang Gu}\email[Corresponding author: ]{qgu@ustb.edu.cn}
\affiliation{Department of Physics, University of Science and
Technology Beijing, Beijing 100083, China}

\date{\today}

\begin{abstract}

The charged Fermi gas with a small Lande-factor $g$ is expected to
be diamagnetic, while that with a larger $g$ could be paramagnetic.
We calculate the critical value of the $g$-factor which separates
the dia- and para-magnetic regions. In the weak-field limit, $g_{c}$
has the same value both at high and low temperatures,
$g_{c}=1/\sqrt{12}$. Nevertheless, $g_{c}$ increases with the
temperature reducing in finite magnetic fields. We also compare the
$g_{c}$ value of Fermi gases with those of Boltzmann and Bose gases,
supposing the particle has three Zeeman levels $\sigma=\pm1, 0$, and
find that $g_{c}$ of Bose and Fermi gases is larger and smaller than
that of Boltzmann gases, respectively.
\end{abstract}


\maketitle

\section{Introduction}

Magnetism of electron gases has been considerably studied in
condensed matter physics. In magnetic field, the magnetization of a
free electron gas consists of two independent parts. The spin
magnetic moment of electrons results in the paramagnetic part (the
Pauli paramagnetism), while the orbital motion due to charge degree
of freedom in magnetic field induces the diamagnetic part (the
Landau diamagnetism)\cite{Landau}. The Pauli paramagnetism and the
Landau diamagnetism compete with each other. For electrons whose
Lande-factor $g=2$, the zero-field paramagnetic susceptibility is
two times stronger than the diamagnetic susceptibility, so the free
electron gas exhibits paramagnetism altogether.

Magnetic properties of relativistic Fermi gases have also been under
extensive investigation. Daicic {\it et al.} developed statistical
mechanics for the magnetized pair-fermion gases and found that the
intrinsic spin causes important effects upon the relativistic para-
and dia-magnetism\cite{Daicic}.

The study of ultracold atoms has stimulated renewed research
interest in the magnetism of quantum
gases\cite{Kurn,Stenger,Sadler,Jo}. When atomic gases are confined
in optical traps\cite{Kurn,Stenger}, their spin degree of freedom
becomes active, leading to the manifestation of magnetism.
Theoretically, the paramagnetism\cite{Yamada} and
ferromagnetism\cite{Ho,Gu3,Szirmai} in a neutral spin-1 Bose gas
have ever been studied. In experiments, magnetic domains have been
directly observed in $^{87}$Rb condensate, a typical ferromagnetic
spinor condensate\cite{Sadler}. Very recently, the exploration of
magnetism in quantum gases has been extended to the Fermi gas. It is
already observed that an ultracold two-component Fermi gas of
neutral $^{6}\text{Li}$ atoms exhibits ferromagnetism caused by
repulsive interactions between atoms\cite{Jo}.

Furthermore, it is possible to realize charged quantum gases from
neutral ultracold atoms. So far, cold plasma has been created by
photoionization of cold atoms\cite{Killian}. The temperatures of
electrons and ions are as low as 100 mK and 10 $\mu$K, respectively.
The ions can be regarded as charged Bose or Fermi gases. Once the
quantum gas has both the spin and charge degrees of freedom, there
arises the competition between paramagnetism and diamagnetism, as in
electrons. Different from electrons, the $g$-factor for different
magnetic ions is diverse, ranging from 0 to 2.

The $g$-factor is a characteristic parameter to measure the
intensity of paramagnetic effect. It is expected that the quantum
gas displays diamagnetism in small $g$ region, but paramagnetism in
large $g$ region. The main purpose of this paper is to calculate the
critical value of $g$. We also present a comparison with results of
charged spin-1 Bose gases which have been obtained
previously\cite{Jian}.

\section{The Model}

We consider a charged Fermi gas of $N$ particles, with the effective
Hamiltonian written as
\begin{eqnarray}\label{Hamil}
\bar{H} - \mu {N}= D_L\sum_{j,k_z,\sigma}\left(\epsilon_{jk_{z}}
+\epsilon_{\sigma} - \mu\right)n_{jk_z\sigma},
\end{eqnarray}
where $\mu$ is the chemical potential and $\epsilon_{jk_{z}}$ is the
quantized Landau energy in the magnetic field $B$,
\begin{eqnarray}\label{Diam}
\epsilon_{jk_{z}}=(\frac{1}{2}+j)\hbar\omega + \frac{\hbar^{2}
k_{z}^{2}}{2m^{\ast}}
\end{eqnarray}
with $j=0,1,2,\ldots$ labeling different Landau levels and $\omega
=qB/(m^{\ast}c)$ being the gyromagnetic frequency of a fermion with
charge $q$ and mass $m^*$. $D_L={qBS_{xy}}/({2\pi\hbar c})$ marks
the degeneracy of each Landau level with $S_{xy}$ the total section
area in x, y directions of the system. $\epsilon_{\sigma}$ denotes
the Zeeman energy,
\begin{eqnarray}\label{Param}
\epsilon_{\sigma} = - g\frac{\hbar q}{m^*c} \sigma B
  = -g\sigma \hbar\omega,
\end{eqnarray}
where $g$ is the Lande-factor and $\sigma$ refers to the spin-$z$
index of Zeeman state $\left| {F,\sigma} \right\rangle$.

The grand thermodynamic potential of the Fermi gas is expressed as
\begin{align}\label{T1}
\Omega_{T\neq0}& =-\frac{1}{\beta}D_L\sum_{j,k_z,\sigma}\ln
[1+e^{-\beta(\epsilon_{jk_{z}} +\epsilon_{\sigma} - \mu )}],
\end{align}
where $\beta=(k_{B}T)^{-1}$. Performing Taylor expansions and then
integrating out $k_z$, Eq.~(\ref{T1}) becomes
\begin{align}\label{T2}
\Omega_{T\neq0}=&-\frac{\omega
V}{\hbar^{2}}\left(\frac{m^{\ast}}{2\pi\beta}\right)^{3/2}\nonumber\\
&\times\sum_{l=1}^{\infty}\sum_{\sigma}\frac{(-1)^{l+1}l^{-\frac{3}{2}}e^{-l\beta(\frac{\hbar\omega}{2}
 -g\sigma\hbar\omega- \mu)}}{1-e^{-l\beta\hbar\omega}},
\end{align}
where $V$ is the volume of the system. For simplicity, the following
notation is introduced,
\begin{align}\label{s}
F_{\tau}^{\sigma}[\alpha,\delta]=\sum_{l=1}^{\infty}\frac{(-1)^{l+1}l^{\alpha/2}e^{-lx(\eta_{\sigma}+\delta)}}{(1-e^{-lx})^{\tau}},
\end{align}
where $x=\beta\hbar\omega$ and $\eta_{\sigma}=(\hbar\omega/2-
\mu+\epsilon_{\sigma})/(\hbar\omega)$. Then Eq.~(\ref{T2}) is
rewritten as
\begin{align}\label{Ts}
\Omega_{T\neq0}=-\frac{\omega
V}{\hbar^{2}}\left(\frac{m^{\ast}}{2\pi\beta}\right)^{3/2}\sum_{\sigma}F_{1}^{\sigma}[-3,0].
\end{align}
For a system with the given particle density $n=N/V$, the chemical
potential is obtained according to the following equation
\begin{align}\label{n}
n = -\frac{1}{V}\left(\frac{\partial\Omega}{\partial\mu}\right)_{T,V}
  = x\left(\frac{m^{\ast}}{2\pi\beta\hbar^{2}}\right)^{3/2}\sum_{\sigma}F_{1}^{\sigma}[-1,0].
\end{align}
A similar treatment has been employed to study diamagnetism of the
charged spinless Bose gas\cite{Toms} and extended to the study of
competition of diamagnetism and paramagnetism in charged spin-1 Bose
gases\cite{Jian}. This method is more applicable at relatively high
temperatures.

To determine $g_c$, we need calculate the magnetization density $M$
as a function of the magnetic field $B$ and temperature $T$. The
system is paramagnetic when $M>0$ while diamagnetic when $M<0$. $M$
can be derived from the thermodynamic potential by the standard
procedure $M_{T\neq 0} = -
\frac{1}{V}\left(\frac{\partial\Omega}{\partial B}\right)_{T,V}$,
which yields
\begin{align}\label{Mag}
M_{T\neq 0} &=\frac{\hbar q}{m^{\ast}c}\left(\frac{m^{\ast}}{2\pi\beta\hbar^{2}}\right)^{3/2}\sum_{\sigma}\biggl\{F_{1}^{\sigma}[-3,0] \nonumber\\
&+x(g\sigma-\frac{1}{2})
F_{1}^{\sigma}[-1,0]-xF_{2}^{\sigma}[-1,1]\biggr\}.
\end{align}

For carrying out numerical calculations, it is useful to make the
parameters dimensionless, such as $\overline{M}=m^{\ast}cM/(n\hbar
q)$, $\overline{\omega}=\hbar\omega/(k_{B}T^{\ast})$, and
$t=T/T^{\ast}$. Here $T^{\ast}$ is the characteristic temperature of
the system, which is given by $k_{B}T^{\ast}=2\pi\hbar^{2}
n^{\frac{2}{3}}/m^{\ast}$. Then we have $x=\overline{\omega}/t$,
$\eta_{\sigma}=1/2-\mu^{\prime}/\overline{\omega}-g\sigma $ and
$\mu^{\prime}=\mu/(k_{B}T^{\ast})$ in the notation
$F_{\tau}^{\sigma}[\alpha,\delta]$. Equations (\ref{n}) and
(\ref{Mag}) can be re-expressed as
\begin{align}\label{n1}
1=\overline{\omega}t^{1/2}\sum_{\sigma}F_{1}^{\sigma}[-1,0]
\end{align}
and
\begin{align}\label{M1}
\overline{M}_{T\neq 0}
=& t^{3/2}\sum_{\sigma}\biggl\{F_{1}^{\sigma}[-3,0]+x(g\sigma-\frac{1}{2})\nonumber\\
&\times F_{1}^{\sigma}[-1,0]-xF_{2}^{\sigma}[-1,1]\biggr\}.
\end{align}
In following discussions, $\overline{\omega}$ is used to indicate
the magnitude of magnetic field  since it is proportional to $B$.

\section{Results and discussions}

\begin {figure}[t]
\includegraphics[width=0.45\textwidth,keepaspectratio=true]{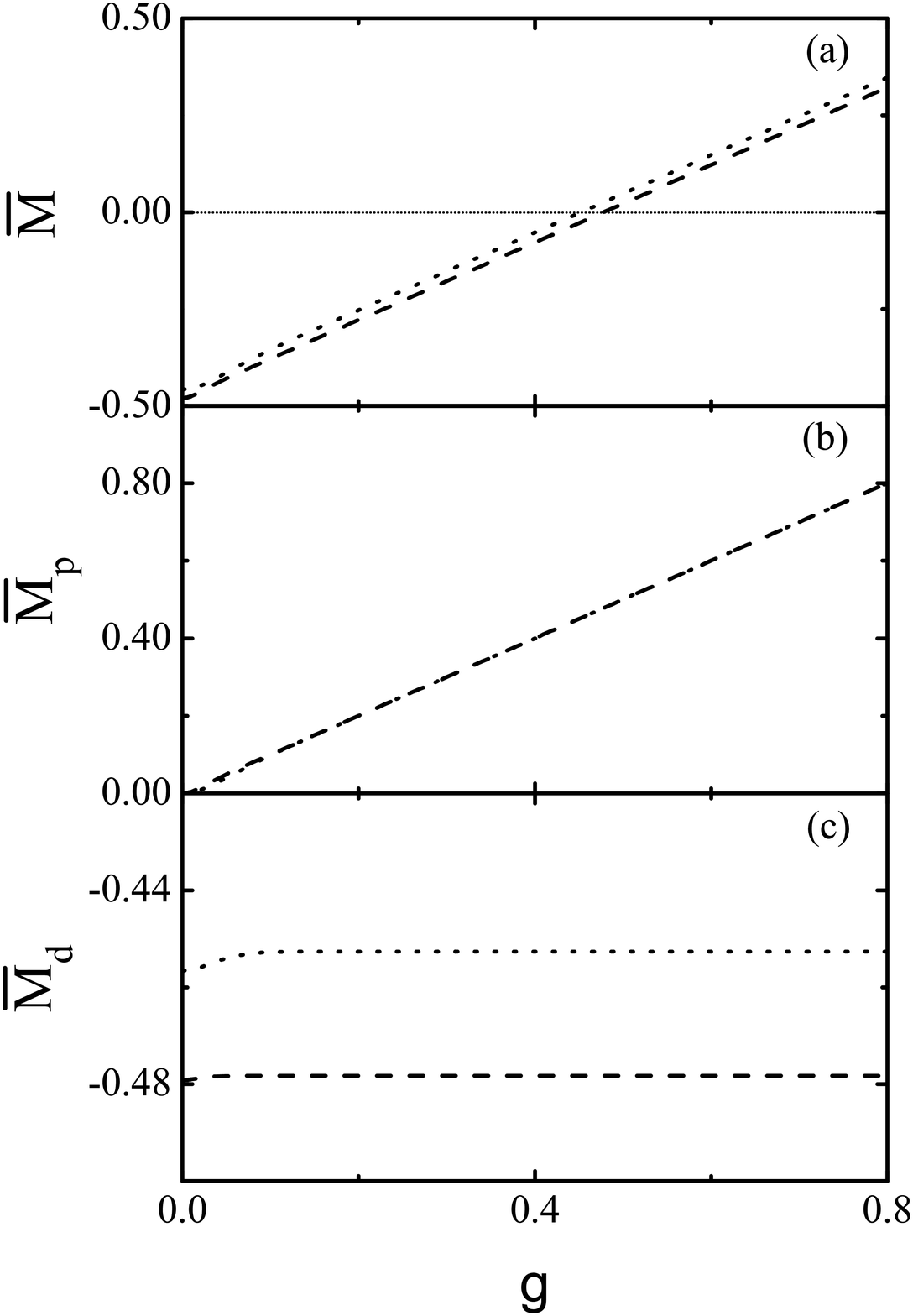}
\caption{(a) The total magnetization density ($\overline{M}$), (b)
the paramagnetization density ($\overline{M}_{p}$), and (c) the
diamagnetization density ($\overline{M}_{d}$) as a function of $g$
at $t=0.2$. The dotted and dashed lines correspond to
$\overline{\omega}= 5$ and $10$, respectively. }\label{M-G}
\end{figure}

First, we look at a spin-$\frac12$ Fermi gas, setting $\sigma=\pm 1$
to present the two Zeeman levels. The dimensionless magnetization
density $\overline{M}$ as a function of $g$ is shown in
Fig.~\ref{M-G}(a). As expected, $\overline{M}$ is negative in the
small $g$ region, which means that the diamagnetism dominates. For
each given value of $\overline{\omega}$, $\overline{M}$ grows with
$g$ and changes its sign from negative to positive in the larger $g$
region, indicating that the paramagnetic effect is enhanced due to
increase of $g$. This phenomenon is also observed in the Bose
system\cite{Jian}. $g_{c}$ is just the value of $g$ where
$\overline{M}=0$.

The total magnetization $\overline{M}$ shown in Fig.~\ref{M-G}(a)
consists of both the paramagnetic and diamagnetic contributions.
Figure~\ref{M-G}(b) depicts the pure paramagnetic contribution to
$\overline{M}$, which is calculated by $\overline{M}_{p}= g m$ where
$m= n_{1}-n_{-1}$. Figure \ref{M-G}(c) plots the diamagnetic
contribution to $\overline{M}$, $\overline{M}_{d}=
\overline{M}-\overline{M}_p$. For each fixed value of
$\overline{\omega}$, the diamagnetization $\overline{M}_{d}$ is
slightly weakened with increasing $g$. Interestingly, for a charged
spin-1 Bose gas in a constant magnetic field, the diamagnetism is
not suppressed but enhanced as $g$ becomes larger\cite{Jian}.
Comparing Figs \ref{M-G}(a), \ref{M-G}(b) and \ref{M-G}(c), it can
be seen clearly that the increase of $\overline{M}$ with $g$ is
mainly owing to the paramagnetization $\overline{M}_{p}$.

\begin {figure}[t]
\includegraphics[width=0.45\textwidth,keepaspectratio=true]{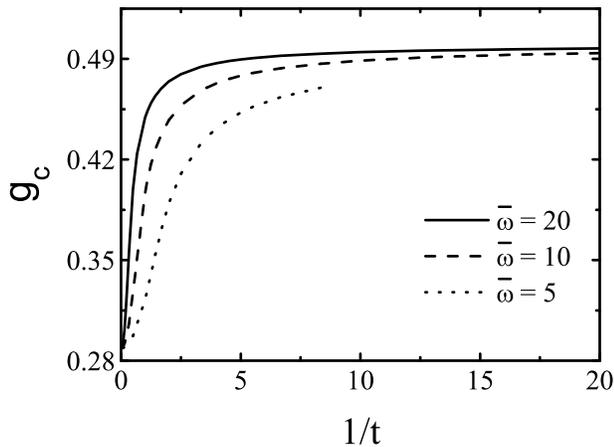}
\caption{Plots of the critical value of $g$-factor as a function of
$1/t$. The magnetic field is chosen as $\overline{\omega}$ = $20$
(solid line), $10$ (dashed line) and $5$ (dotted line),
respectively.} \label{Gc}
\end{figure}

$g_{c}$ is an important parameter to describe the competition
between the diamagnetism and paramagnetism. Figure~\ref{Gc} plots
$g_{c}$ for the charged spin-$\frac 12$ Fermi gas. The temperature
is described by $1/t$ and the magnetic fields are chosen to be
relatively larger, since the method adopted here is more applicable
at higher temperatures or in stronger fields. In the high
temperature limit, $g_{c}$ tends to a universal value,
$g_{c}|_{t\to\infty}= 0.28868$, regardless of the strength of the
magnetic field. In a fixed magnetic field, $g_{c}$ increases
monotonically as the temperature falls down. But the trend of
increasing is slowed down if the magnetic field is weakened.

Present method can not produce valid results in the low temperature
region if a small magnetic field is chosen. In this case that
$g\overline{\omega}\ll t\ll \mu^{\prime}$, the grand thermodynamic
potential in Eq.~(\ref{T1}) can be calculated using the
Euler-Maclaurin formula
\begin{align}\label{l}
\sum_{j=0}^{\infty}\psi(j+\frac{1}{2})\approx
\int_{0}^{\infty}\psi(x)dx+\frac{1}{24}\psi'(0),
\end{align}
and then Eq.~(\ref{T1}) is transformed into
\begin{align}\label{ol}
\Omega_{T\neq0}& =-\frac{V}{\beta\lambda^{3}}
\sum_{\sigma}f_{\frac{5}{2}}(z)+\frac{V\beta}{24\lambda^{3}}(\hbar\omega)^{2}\sum_{\sigma}f_{\frac{1}{2}}(z),
\end{align}
where $\lambda=h\beta^{\frac{1}{2}}/(2\pi m^{\ast})^{\frac{1}{2}}$,
$z=e^{\beta(\mu-\epsilon_{\sigma})}$ and the Fermi-Dirac integral
$f_{n}(z)$ is normally defined as
\begin{align}\label{f}
f_{n}(z)& \equiv
\frac{1}{\Gamma(n)}\int_{0}^{\infty}\frac{x^{n-1}}{z^{-1}e^{x}+1}dx,
\end{align}
where $\Gamma(n)$ is a usual gamma function and $x=\beta\epsilon$.
Then the magnetization density, $\overline{M}$, is obtained from
grand thermodynamic potential in Eq.~(\ref{ol}),
\begin{align}\label{Ml}
\overline{M}_{T\neq 0} =&gt^{3/2}\sum_{\sigma}\sigma
f_{\frac{3}{2}}(z^{\prime})-\frac{\overline{\omega}t^{1/2}}{12}\sum_{\sigma}f_{\frac{1}{2}}(z^{\prime})\nonumber\\
&-g\frac{\overline{\omega}^{2}t^{-1/2}}{24}\sum_{\sigma}\sigma
f_{-\frac{1}{2}}(z^{\prime}),
\end{align}
where $z^{\prime}=e^{(\mu^{\prime}+g\sigma\overline{\omega})/t}$.
After some algebra, we get from Eq.~(\ref{Ml}) that
$g_{c}|_{\overline{\omega}\to0}= 1/\sqrt{12}\approx0.28868$. So in
the weak field limit, $g_c$ has the same value both at the high and
low temperature limit. This implies that the $g_{c}-1/t$ curve is
likely to flatten out in the weak-field limit.

\begin {figure}[t]
\includegraphics[width=0.45\textwidth,keepaspectratio=true]{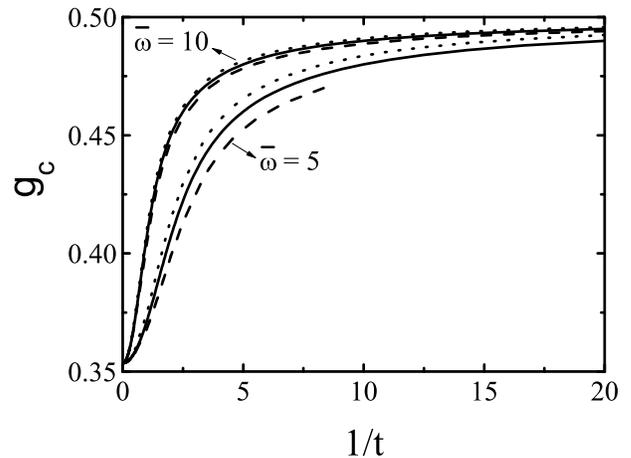}
\caption{The $g_{c}- 1/t$ curves for charged spin-1 gases obeying
the Bose-Einstein (BE, dotted line), Maxwell-Boltzmann (MB, solid
line) and Fermi-Dirac (FD, dashed line) statistics, respectively.
The magnetic field is chosen as $\overline{\omega}$ = $10$ and $5$.}
\label{Gc-Bz}
\end{figure}

To proceed, we make a comparison between Fermi and Bose gases.
Considering that $g_c$ for charged spin-1 Bose gas has already been
studied\cite{Jian}, we discuss a Fermi gas with three sublevels,
$\sigma=\pm1,0$ (a pseudo-spin-1 Fermi gas) to ensure the
comparability. Results for the spin-1 Boltzmann gas are also
obtained. Figure \ref{Gc-Bz} shows the $g_{c}-1/t$ curves for the
three kinds of charged spin-1 gases. In a given value of
$\overline{\omega}$, $g_{c}$ of all the three gases displays similar
temperature-dependence: $g_{c}$ increases monotonously as $t$
reduces. In the high temperature limit, $g_{c}$ goes to the same
value in all magnetic fields, $g_{c}|_{t\to\infty}=1/\sqrt{8}$,
reflecting that the Bose-Einstein (BE) and Fermi-Dirac (FD)
statistics coincide with the Maxwell-Boltzmann (MB) statistics in
this case.

Figure \ref{Gc-Bz} also demonstrates the difference among the three
kinds of statistics. For each given fixed value of
$\overline{\omega}$, the $g_{c}-1/t$ curves of Bose and Fermi gases
always locate at the two sides of that of the Boltzmann gas. Given
the same temperature and magnetic field, $g_{c}$ of Fermi gas is the
smallest. According to our previous research\cite{Jian},
$g_{c}|_{t\to0}=1/2$ for the Boltzmann gas regardless of the
magnetic field. This means that $g_{c}$ of the Fermi gas does never
exceed $1/2$ in the low temperature, no matter how strong the field
is.

\section{Summary}

In summary, we study the interplay between paramagnetism and
diamagnetism of charged Fermi gases supposing that the Lande-factor
$g$ is a variable. The gas undergoes a shift from diamagnetism to
paramagnetism at the critical value of $g$ and $g_{c}$ increases
monotonically as the temperature $t$ decreases in a fixed magnetic
field $\overline{\omega}$, and the rise in $g_{c}$ is lowered as
$\overline{\omega}$ is reduced. We conjecture that $g_{c}$ holds a
constant at all temperatures in the weak field limit. For a
spin-$\frac12$ Fermi gas,
$g_{c}|_{\overline{\omega}\to0}=1/\sqrt{12}$. We also briefly
compare $g_{c}$ of charged spin-1 gases obeying the Fermi-Dirac,
Bose-Einstein and Maxwell-Boltzmann statistics. The $g_{c}-1/t$
curves of Boltzmann gases are always between those of Bose and Fermi
gases in the same magnetic field. In the high temperature limit,
$g_{c}$ of all the three gases tends to the same value.

This work was supported by the Fok Ying Tung Education Foundation of
China (No.~101008), the Key Project of the Chinese Ministry of
Education (No.~109011), and the Fundamental Research Funds for the
Central Universities of China.

\end{document}